\documentclass[aps,pra,superscriptaddress,twocolumn]{revtex4-1}
\usepackage{dcolumn}
\usepackage{amsmath}
\usepackage{amsfonts}
\usepackage{amssymb}
\usepackage{graphicx}
\usepackage{hyperref}
\usepackage{xcolor}

\begin{document}
\title{Evaluation of Twisted Gaussian Schell Model beams produced with phase randomized coherent fields}
\author{G.~Ca\~{n}as}
\affiliation{Departamento de F\'isica, Universidad del B\'io-B\'io, Collao 1202, 5-C Concepci\'on, Chile}
\affiliation{Millennium Institute for Research in Optics, Universidad de Concepci\'on, 160-C Concepci\'on, Chile}
\author{E. S. G\'omez}
\affiliation{Departamento de F\'{\i}sica, Universidad de Concepci\'on, 160-C Concepci\'on, Chile}
\affiliation{Millennium Institute for Research in Optics, Universidad de Concepci\'on, 160-C Concepci\'on, Chile}
\author{G. H. dos Santos}
\affiliation{Departamento de F\'{i}sica, Universidade Federal de Santa Catarina, CEP 88040-900, Florian\'{o}plis, SC, Brazil}
\author{A. G. de Oliveira}
\affiliation{Departamento de F\'{i}sica, Universidade Federal de Santa Catarina, CEP 88040-900, Florian\'{o}plis, SC, Brazil}
\author{N. Rubiano da Silva}
\affiliation{Departamento de F\'{i}sica, Universidade Federal de Santa Catarina, CEP 88040-900, Florian\'{o}plis, SC, Brazil}
\author{Stuti Joshi}
\affiliation{Optics and Photonics Centre, Indian Institute of Technology Delhi, Hauz Khas, New Delhi 110016, India}
\author{Yaseera Ismail}
\affiliation{Quantum Research Group, School of Chemistry and Physics, University of KwaZulu-Natal, Durban 4001, South Africa}
\author{P. H. Souto Ribeiro}
\email{p.h.s.ribeiro@ufsc.br}
\affiliation{Departamento de F\'{i}sica, Universidade Federal de Santa Catarina, CEP 88040-900, Florian\'{o}plis, SC, Brazil}
\author{S. P. Walborn}
\email{swalborn@udec.cl}
\affiliation{Millennium Institute for Research in Optics, Universidad de Concepci\'on, 160-C Concepci\'on, Chile}
\affiliation{Departamento de F\'{\i}sica, Universidad de Concepci\'on, 160-C Concepci\'on, Chile}

 \begin{abstract}
 The twisted Gaussian Schell Model describes a family of partially coherent beams that present several interesting characteristics, and as such have attracted attention in classical and quantum optics.  Recent techniques have been demonstrated to synthesize these beams from a coherent source using a discrete set of ``pseudo-modes", where the phase of each mode is randomized so that they are mutually incoherent.   Here we investigate this technique and evaluate the resulting beam parameters, such as divergence, coherence length and twist phase.  We show that for a finite set of modes there is also some residual coherence, which can have an observable effect. A theoretical model is developed for the output field that includes residual coherence and agrees very well with experimental data. In addition, we demonstrate a simple method to measure the twist phase using double slit interference.  
 \end{abstract}

\pacs{05.45.Yv, 03.75.Lm, 42.65.Tg}
\maketitle

\section{Introduction}
Spatial coherence is one of the fundamental properties of a light field that describes the correlation between fluctuating electric field components at two spatial points. Optical beams with low spatial coherence, such as Gaussian Schell-model (GSM) beams, are extensively studied due to their wide range of applications in imaging, free-space optical communication, optical scattering, nonlinear optics, etc \cite{Cai2022,ismail2017,ismail2020,Hutter21,Zhang:19,Ma:17}. Simon and Mukunda have introduced a position dependent twist phase in the correlation function of a GSM beam \cite{simon93} which was first experimentally realized by Friberg et. al. \cite{friberg94}.
The family of partially coherent beams that possess twist-phase are termed as twisted Gaussian Schell-model (TGSM) beams. The twist phase is not an optical  phase in the usual sense, as its modulus is bounded by the inverse of the square of the spatial coherence length, therefore, it can only exist when the beams are partially coherent. The twist phase is related to orbital angular momentum of the beam, and is responsible for the rotation of the beam along the propagation direction. The sense of rotation can be defined by the handedness of the beam.  Additionally, other classes of TGSM beams such as twisted Laguerre Gaussian Schell-model (TLGSM) \cite{Peng:18}, twisted Hermite GSM (THGSM) beams \cite{Xiaofeng2022}, ring-shaped twisted Gaussian Schell-model array (RTGSMA) \cite{Zheng2020}, and twisted vortex Gaussian Schell-model (TVGSM) beams \cite{Stahl:18} have been introduced and their propagation properties have been studied. Due to the unique properties of the twist phase, TGSM beams find applications in various research areas such as optical communication through atmospheric and underwater turbulence \cite{Cai2006,Wang:10,Wang:12,Peng:17,Zhou:18}, in resisting coherence induced depolarization, overcoming the classical Rayleigh limit \cite{Tong:12}, to control the coherence of optical solitons \cite{Ponomarenko2001}, to boost entanglement in photon pairs \cite{Hutter20}, and in stimulated parametric down-conversion \cite{dos2022phase}.

Despite the extensive theoretical progress in studies involving TGSM beams, and the appeal of these beams in several applications, very few experimental attempts have been made to generate, characterize, and study their propagation properties \cite{friberg94,wang19,Stahl:18,tian20,zhang2021generating,Liu:22, wang2022experimental}. The experimental setup used in \cite{friberg94} consisted of a complex optical system which was the combination of six-cylindrical lenses and a variable-coherence anisotropic GSM source. A different method was used by Wang et. al., which have developed a method to generate twisted Gaussian Schell-model (TGSM) beams by converting an anisotropic GSM beam into a TGSM beam with a set of three cylindrical lenses \cite{wang19}.
The generation of TGSM beams has also been demonstrated by implementing the continuous coherent beam integral function in a discrete form \cite{tian20}. More recently, the generation of TGSM beams with controllable twist phase using an incoherent superposition of random modes obeying Gaussian statistics has been reported \cite{wang2022experimental}. Moreover, besides the usual TGSM beams, an alternative kind of partially coherent vector beam named {\em radially polarized twisted partially coherent vortex}  (RPTPCV) beam was also generated \cite{Liu:22}. It was demonstrated that the twist phase, vortex phase, polarization and coherence all together influence the far-field statistical properties of the RPTPCV beam.\\

In this paper we study the TGSM beam generation method demonstrated in Ref. \cite{tian20}, which uses a finite sequence of images displayed on a spatial light modulator.  Since this method uses a discretized form of a decomposition into a continuous set of non-orthogonal ``pseudo-modes", it is an approximation of a TGSM beam. Here we implement this technique and characterize the parameters of the beams that are produced. In particular, we observe corrections to the transverse coherence length that arise from the coherent background light.  Previous work has shown results for only one or two values of the twist phase and/or coherence length.  Here we show results for a wide range of values, and explore several known techniques for evaluating twist phase and coherence length. We also propose a new method to measure the twist phase using double slit interference, so that this type of setup can be used to obtain the coherence length and twist phase from a single 2D interferogram.  We expect these results to be very useful in performing experiments where a fine control of the coherence length and twist phase is required.
 \section{Theory}
 \label{sec:theory}
 The cross spectral density (CSD) of a monochromatic scalar field can be decomposed as a convex combination given by \cite{Born64}
 \begin{equation}
 \Gamma(\mathbf{r}_1,\mathbf{r}_2) = \iint d\mathbf{v} p(\mathbf{v})  K^*(\mathbf{r}_1,\mathbf{v})  K(\mathbf{r}_2,\mathbf{v}), 
 \label{eq:wint}
 \end{equation}
 where $K$ are non-orthogonal modes or ``pseudo-modes" \cite{wang20}. Here $\mathbf{r}_j=(x_j,y_j)$ and $\mathbf{v}=(v_x,v_y)$ are two-dimensional vectors in the transverse plane. The function $p(\mathbf{v})$ is a weight function. This pseudo-mode decomposition is quite useful as it can be used to synthesize partially coherent beams \cite{wang20,tian20,wang21}. 
 \par
 A TGSM beam can be described by 
 \begin{equation}
 \Gamma_{TG}(\mathbf{r}_1,\mathbf{r}_2) = e^{-\frac{r_1^2+r_2^2}{4 \sigma^2}}e^{-\frac{|\mathbf{r}_1-\mathbf{r}_2|^2}{2 \delta^2}}e^{-i k \mu(x_1 y_2-y_1x_2)}.
 \label{eq:TGSM}
 \end{equation}
 where $k$ is the wave number and $\sigma$ is the beam waist. The parameter $\mu$ is the twist phase, such that $|\mu| \leq 1/(k \delta^2)$ , where $\delta$ is the transverse coherence length. One can also define the normalized twist phase $\tau= k \delta^2 \mu$. 
 \par
 To express the CSD \eqref{eq:TGSM} in the form \eqref{eq:wint}, we follow Refs. \cite{wang20,tian20}, and find that the mode functions 
 \begin{align}
 \label{eq:K}
 K(\mathbf{r},\mathbf{v}) = & \exp \left [-\frac{\sigma^2}{2 a \sigma^2+1}\left (\frac{\mathbf{r}}{2 \sigma^2} + a \mathbf{r} - a \mathbf{v} \right)^2  \right]  
 \nonumber \\
 & \times \exp \left [ -i k \mu(x v_y - y v_x) \right] 
 \end{align}
 and weight functions
 \begin{equation}
 p(\mathbf{v})  =\exp \left(-\frac{av^2}{ \left(2 a \sigma^2+ 1\right)}\right),
 \end{equation}
with parameter
 \begin{equation}
 a=\frac{1}{\delta^2} \left(1+\sqrt{1-k^2\mu^2\delta^2}\right),  \label{eq:a}
 \end{equation}
 when plugged into \eqref{eq:wint} and integrated, result in a TGSM beam with cross spectral density given by \eqref{eq:TGSM}. 
 
 \par
 In practice, to construct an arbitrary partially coherent field from an input coherent field a finite set of modes $K$ is used.  Thus, one needs to replace the integral in \eqref{eq:wint} with a finite sum, giving 
 \begin{equation}
 \Gamma(\mathbf{r}_1,\mathbf{r}_2)  \approx  E^*(\mathbf{r}_1) E(\mathbf{r}_2)\sum_{n}^{N} p(\mathbf{v}_n)  K^*(\mathbf{r}_1,\mathbf{v}_{n})  K(\mathbf{r}_2,\mathbf{v}_{n}),
 \label{eq:wsum}
 \end{equation}
 where $ E(\mathbf{r})$ is the optical field illuminating the device. Our goal here is to make the CSD \eqref{eq:wsum} as close as possible to the TGSM beam described by \eqref{eq:TGSM}.  Ideally, the input illuminating field approximates a plane wave such that $ E(\mathbf{r})\sim E_0$ is constant. To determine the number of modes $N$ required for an accurate representation of the partially coherent field given in \eqref{eq:wint}, Refs. \cite{wang20,tian20} have used the degree of coherence (DOC), given by 
 \begin{equation}
\gamma(\mathbf{r}_1,\mathbf{r}_2)= \sqrt{\frac{\Gamma^*(\mathbf{r}_1,\mathbf{r}_2) \Gamma(\mathbf{r}_1,\mathbf{r}_2) }{\Gamma(\mathbf{r}_1,\mathbf{r}_1) \Gamma(\mathbf{r}_2,\mathbf{r}_2)}}.
 \label{eq:DOC}
 \end{equation}
 Calculating the DOC using the exact expression for the TGSM beam \eqref{eq:TGSM} gives $\exp(-|\mathbf{r}_1-\mathbf{r}_2|^2/2 \delta^2)$.  On the other hand, by plugging Eq. \eqref{eq:wsum} into \eqref{eq:DOC}, one can determine the DOC for the approximate field.  Numerical results show that when $N$ is large enough, on the order of a few hundred modes, the exact result can be reproduced with large precision.  
 \par
 To produce the incoherent sum of modes in \eqref{eq:wsum} using a coherent light source, Refs. \cite{wang20,tian20,wang21} have introduced a method where a spatial light modulator (SLM) is used to modulate the amplitude and phase of an input field through a film composed of $L$ images.  Below we will describe the SLM technique in more detail. For now, it suffices to consider that each image is associated to the function
 \begin{equation}
 {\Phi}_l(\mathbf{r}) = \sum_{n=1}^{N} \sqrt{p(\mathbf{v}_n)} K(\mathbf{r},\mathbf{v}_n) e^{i \varphi_{l,n}},  
 \label{eq:Phi}
 \end{equation}
 where $l$ is the image index running from 1 to $L$. The phases $\varphi_{l,n}$ are randomly chosen between $0$ and $2 \pi$ for each mode in each image, while the displacement vectors $\mathbf{v}_n=(v_{nx},v_{ny})$ are chosen uniformly within a sub-area of the SLM. In this way, the cross spectral density of the output field averaged over the $L$ images is 
 \begin{align}
 \Gamma(\mathbf{r}_1,\mathbf{r}_2) = & E^*(\mathbf{r}_1) E(\mathbf{r}_2) \sum_{n,m}^{N} \sqrt{p(\mathbf{v}_n)p(\mathbf{v}_m)} \times  \nonumber \\
 &  K^*(\mathbf{r}_1,\mathbf{v}_{n})  K(\mathbf{r}_2,\mathbf{v}_{m}) \sum_l^L  e^{i (\varphi_{l,m}-\varphi_{l,n})}.
 \label{eq:wsum2}
 \end{align}
 The sum in $l$ is over the randomly chosen phases, and is responsible for the coherence between different SLM images. For $L \rightarrow \infty$, this sum goes to zero, and we have partial coherence as determined by the chosen value of $\delta$.  For finite $L$ this sum gives $L (\delta_{n,m}+  \Delta_L)$, where the real parameter
 \begin{equation}
 \Delta_L = \frac{1}{L} \sum_{l, s.t. n\neq m}^L e^{i (\varphi_{l,m}-\varphi_{l,n})}
 \label{eq:DeltaL}
 \end{equation}
 can be thought of as a residual coherence between the pseudo-modes. Evaluating expression \eqref{eq:DeltaL} for $L$ from $10$ to $19000$ with $1000$ random samples for each data point, we obtain the mean $\langle \Delta_L\rangle$ and standard deviation $\sigma_{\Delta_L}$. By curve fitting, we find that the mean values are well described by the expression $\langle \Delta_L\rangle \approx 3.56(\exp(1/4 \sqrt{L})-1)$, with standard deviation $\sigma_{\Delta_L} \approx \langle \Delta_L\rangle /2$.  For $L=300$, our simulation gives $\Delta_{300} \approx0.051 \pm 0.027$.  For $L=19000$, we find $\Delta_{19000} \approx0.0064 \pm 0.0033$. To achieve the ideal case $\Delta_L= 0$, a very large number of images is required. However, this can lead to very long sampling times, since the refresh rate of SLMs and the frame rate of CCD cameras and similar devices are typically on the order of tens of Hz. Thus, in most applications, which are limited to sampling around a few hundred images, a coherent background is present, and might have noticeable consequences as we will demonstrate in the following sections.  
 \par
 Another experimental parameter that can have relevant consequences is the field used to illuminate the SLM. We assume that this is a coherent beam with a Gaussian spatial profile, given by
 \begin{equation}
     E(\mathbf{r}) = E_0 e^{-r^2/4 w^2}e^{ -i k r^2/2 R},
 \end{equation}
 where $2 w$ is the beam waist and $R$ is the radius of phase curvature.  To achieve the ideal case of plane wave illumination, these parameters should be much sufficiently large so that the amplitude and phase profile of the illuminating field can be considered to be constant. We will see in the following section that both $w$ and $R$ can have relative consequences on the parameters of the synthesized beam. 
 \par
 To take these issues into account, we return to the output CSD \eqref{eq:wsum2}, and notice from Eq. \eqref{eq:DeltaL} that the output field can then be written as an incoherent combination of the desired TGSM beam, together with a coherent background field
 \begin{equation}
 \Gamma(\mathbf{r}_1,\mathbf{r}_2) =  (1-\Delta_L) \Gamma_{TG}(\mathbf{r}_1,\mathbf{r}_2) + \Delta_L \Gamma_{coh}(\mathbf{r}_1,\mathbf{r}_2),
 \label{eq:Wtot}
 \end{equation}
 with the CSD of the the TGSM beam given by Eq. \eqref{eq:TGSM}, the CSD of the coherent field given by
 \begin{align}
 \Gamma_{coh}(\mathbf{r}_1,\mathbf{r}_2) = &  E^*(\mathbf{r}_1) \sum_{n}^{N} \sqrt{p(\mathbf{v}_n)}  K^*(\mathbf{r}_1,\mathbf{v}_{n})  \times  \nonumber \\
 &E(\mathbf{r}_2) \sum_{m}^{N} \sqrt{p(\mathbf{v}_m)}   K(\mathbf{r}_2,\mathbf{v}_{m})
 \label{eq:Wcoh}
 \end{align}
 and relative weights given by $1-\Delta_L$ and $\Delta_L$, respectively. Here we have also dropped a multiplicative factor $L$ for convenience. 
In the next section, we explore this model and the role of the background coherent field experimentally.
 
 \section{Experiment}

 To synthesize the partially coherent TGSM beams, we produce films of 300 grayscale images, where each image is composed of $N=23 \times 23=529$ Gaussians $K$ given by Eq. \eqref{eq:K}. Again following Refs. \cite{wang20,tian20}, we choose the components of the vector $\mathbf{v}_n=(v_{nx},v_{ny})$ uniformly in a $23\times23$ grid within a range defined by twice the waist of $\sqrt{p(\mathbf{v}_n)}$ where the weight function is appreciable, corresponding to the interval $[-2 \sqrt{(2 \sigma^2+1/a)},2 \sqrt{(2 \sigma^2+1/a)}]$. We use the first-order diffraction of the SLM. To do so, the images are constructed by first defining a uniform phase grating $\propto 2 \pi u_0 X$ modulo $2 \pi$, with first-order diffraction angle determined by the spatial angular frequency $2 \pi u_0$, and $X$ being the horizontal coordinate on the SLM.  Superposed on top of this is the sum of mode functions given in Eq. \eqref{eq:Phi}.  

 \begin{figure}
 \includegraphics[width=8.5cm]{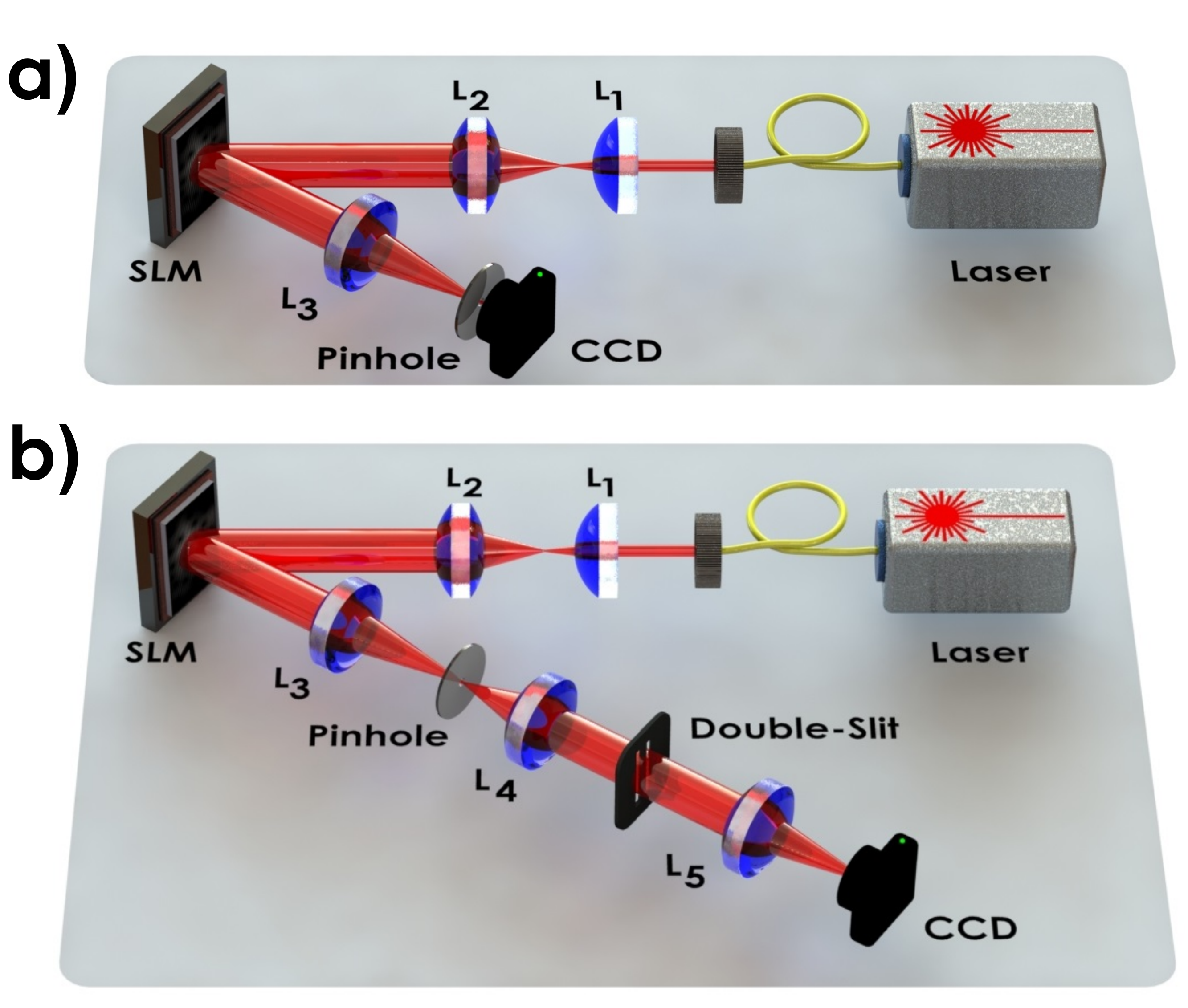}
 \caption{a) Experimental setup for measuring the beam width in the far-field. b) Setup for measurements of coherence and twist phase using a Young double-slit.}
 \label{fig:setup}
 \end{figure}
 \par
 To evaluate the parameters of the TGSM beams, we resort to the setup depicted in Fig \ref{fig:setup}. The source is a continuous-wave (CW) laser, operating at $1550$ nm. The output of the CW laser is connected to a single-mode fiber to obtain a TEM$_{00}$ mode, and then it is expanded via a $4$f optical system (L$_1$ and L$_2$ lenses) so that the entire SLM screen is illuminated. A half-wave plate and polarizing beam splitter (not shown in Fig \ref{fig:setup}) placed at the output of the $4$f optical system are used to set the horizontal polarization of the CW laser. 
 \par
 Then the phase-only SLM (Holoeye Pluto-Telco-013) is illuminated by the CW laser, on which a computer-generated hologram (CGH) is addressed to modulate the amplitude and phase of the incident beam. Several strategies have been implemented to encode arbitrary scalar complex fields on a phase-only CGH \cite{Arrizon:03,Bagnoud:04,Arrizon:07,Zhang:14,Clark:16}. Here, we implement the method proposed by Arrizon et al. \cite{Arrizon:07} to generate type 3 phase-only CGH. 
 \par
 Based on this method, we generate each of the $L$ phase-only CGHs corresponding to each of the functions represented by Eq. \ref{eq:Phi} for index $l$ running from $1$ to $L$. Then, we create a video file composed of the $L$ images as frames, which is used to generate the TGSM beams for different values of $\delta$ and $\mu$. To complete the generation of the TGSM beams, a lens $L_3$ is used to collect the output light and a pinhole placed at the focal plane filters out unwanted diffraction orders and background noise. A CCD camera is used to capture images at the output plane.
 It is important that the data acquisition time be large enough to adequately capture all $L$ images.  This can be done by using a CCD with exposure time that is larger than the time necessary to display all $L$ images on the SLM. Alternatively, one can stroboscopically capture CCD images, and then compute the integrated image.  We choose the second approach, as it allows us to better control the gain on the CCD without creating excessive saturation.  
 \par
 The films are played in a continuous loop on the SLM at a frame rate of 15 frames per second (fps). The CCD camera is set to 15 fps capture rate, with a shutter window $1 s/15 \approx 66.7$ ms. We thus record a single CCD image for each image displayed on the SLM. The CCD images are then added in post-processing, to obtain the full ``integrated" image.    
 \par
 We use several techniques, as shown in Figs. \ref{fig:setup} a) and b), which will be described in the next section, to evaluate the output field produced with this technique.  
\section{Results}
We evaluate several beam properties using the setups shown in Fig. \ref{fig:setup} and compare them with the theoretically predicted values from section \ref{sec:theory}. 
 \subsection{Beam width in the Near-field}
Based on the theory of the previous section, the near field variance of the TGSM beam \eqref{eq:TGSM} is $\sigma^2$.  By direct calculation, we find that the near-field variance of the coherent background \eqref{eq:Wcoh} is $\sigma^2$, so that the near-field variance of the total field \eqref{eq:Wtot} is also $\sigma^2$.  
 \par
 We program the SLM so that $\sigma_0=1$ mm. We measure the near-field variance of TGSM field by imaging the SLM plane onto the CCD camera using an imaging system composed of spherical lenses with focal lengths $L_3=500$ mm and $L_4=150$ mm, as shown in Fig. \ref{fig:setup} b) (CCD placed at doubles slit plane).  We choose values of the coherence length $\delta$ ranging from $0.4$ mm to $5$ mm and normalized twist phase $\tau=0,1$.  Correcting for the magnification factor, we find that the variances in the $x$ and $y$ have a mean value $0.99\pm 0.06$ mm for $\tau=0$ and $0.96\pm 0.05$ mm for $\tau=1$. These values agree with the theoretical prediction $\sigma_0=1$ mm.  The width of the laser beam incident on the SLM was measured to be $2.7\pm 0.3$mm, which is sufficiently large as to approximate the transverse profile as constant.
  \subsection{Beam width in the Far-field}
 
 \begin{figure}
 \includegraphics[width=8cm]{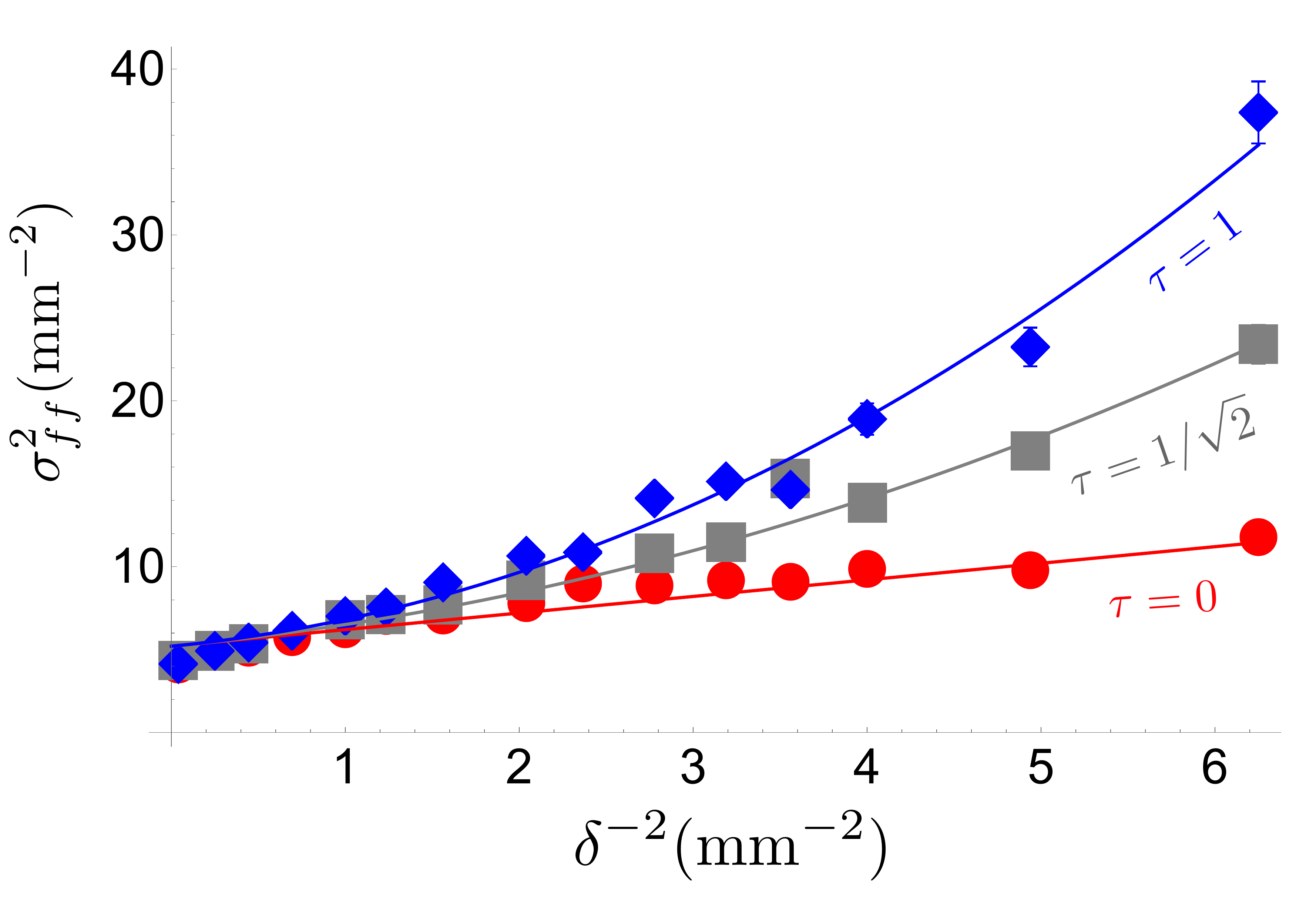}
 \caption{Mean variance in the far field for normalized twist phase $\tau=0, 1/\sqrt{2}, 1$ (from lower to upper curve). The solid curves are theoretical predictions, where the $y$-intercept is an adjustable parameter.}
 \label{fig:farfield}
 \end{figure}
 As a second evaluation method, we analyze the width of the beam in the far field. For the TGSM beam \eqref{eq:TGSM}, the far-field variance is  $1/(4 \sigma^2)+  (1 + k^2 \mu^2 \sigma^2)/\delta^2$, while for the ideal coherent background we use Eq.  \eqref{eq:Wcoh} to calculate $1/(4 \sigma^2)$. However, each of these fields will also acquire the phase curvature of the illuminating beam, resulting in an additional term given by ${k^2\sigma^2}/{R^2}$. For the total field, we then have a far field variance given by
 \begin{equation}
 \label{eq:sigmaff}
 \sigma^2_{ff} = \frac{1}{4\sigma^2} + \frac{k^2\sigma^2}{R^2} +  (1-\Delta_L) \left[ \frac{1}{\delta^2} + \frac{\tau^2 \sigma^2}{\delta^4} \right ],
 \end{equation}
 where we use the definition of the normalized twist phase $\tau$.   Note that the residual coherence factor enters into Eq. \eqref{eq:sigmaff} through the multiplicative factor  $1-\Delta_L$. If sufficient image frames are used, such that $\Delta_L<<1$, we can approximate $1-\Delta_L\sim 1$.
 \par
 Using the experimental setup shown in Fig. \ref{fig:setup} a), we acquire CCD images in the far-field for the entire length of the SLM film.  From the CCD images we calculate the marginal distributions in the $x$ and $y$ directions, and obtain the variance through curve fitting to a Gaussian function. This process was repeated for several values of $\delta$ and normalized twist $\tau=0,1/\sqrt{2},1$. In Fig. \ref{fig:farfield} we show the mean variance (average over $x$ and $y$ directions) as a function of the inverse square of the transverse coherence length $\delta$. The curves are plots of \eqref{eq:sigmaff} using $\sigma$ obtained from the near-field measurements. The correspondence between theory and experiment is quite good. We use the radius of phase curvature $R$ of the laser beam as a free parameter determined by the $y$-intercept of the plots, and find $R \sim 1.45$m, which is reasonable given the optical setup. Thus, we find the divergence of the produced TGSM beams to be in agreement with theory, and close to the ideal value ($\Delta_L=0$).  We note also that the divergence allows one to obtain the square of the twist phase, as was first observed in Ref. \cite{friberg94}.  We also tested the divergence for larger values of $\Delta_L$, where we expect to see larger discrepancies from the ideal case.  However, these results suffer from the fact that the image sequences used to obtain the beams contain only a few frames, which produces integrated images with a transverse profile that is not always Gaussian.  The role of the background coherence will be better evaluated in the following section.     
 
  \subsection{Coherence length}
 The transverse coherence length $\delta$ can be measured directly using double slit interference \cite{fowles89}. Let us consider a double slit aperture in the $x$ direction, with slits at $x=\pm d$, placed in the near-field of a perfect TGSM beam described by \eqref{eq:TGSM}. Integrating over the $y$ degree of freedom, the interference pattern in the far-field of the double slit is
 \begin{equation}
 I_{TG}(x,\delta,\mu) = \gamma_{TG}(d,d) + \mathrm{Re}\left[\gamma_{TG}(d,-d) \right] \cos\left(\frac{2 d k x}{z} \right),
 \label{eq:int1}
 \end{equation}
 where $\gamma_{TG}(d,-d)$ is a shorthand notation for the CSD evaluated at $x_1=d$, $x_2=-d$ and integrated over $y_1=y_2=y$. The visibility $V(\delta,\mu)$ can be calculated, giving
 \begin{equation}
 V(\delta,\mu) = \frac{\mathrm{Re}\left[\gamma_{TG}(d,-d) \right]}{\gamma_{TG}(d,d)} = e^{-2 \frac{d^2}{\delta^2}\left(1+\tau^2 \sigma^2\right)}.
 \label{eq:visTGSM}
 \end{equation}
 We can see that when $\mu$ is known, the visibility is an indicator of $\delta$.
 \begin{figure}
 \includegraphics[width=8.5cm]{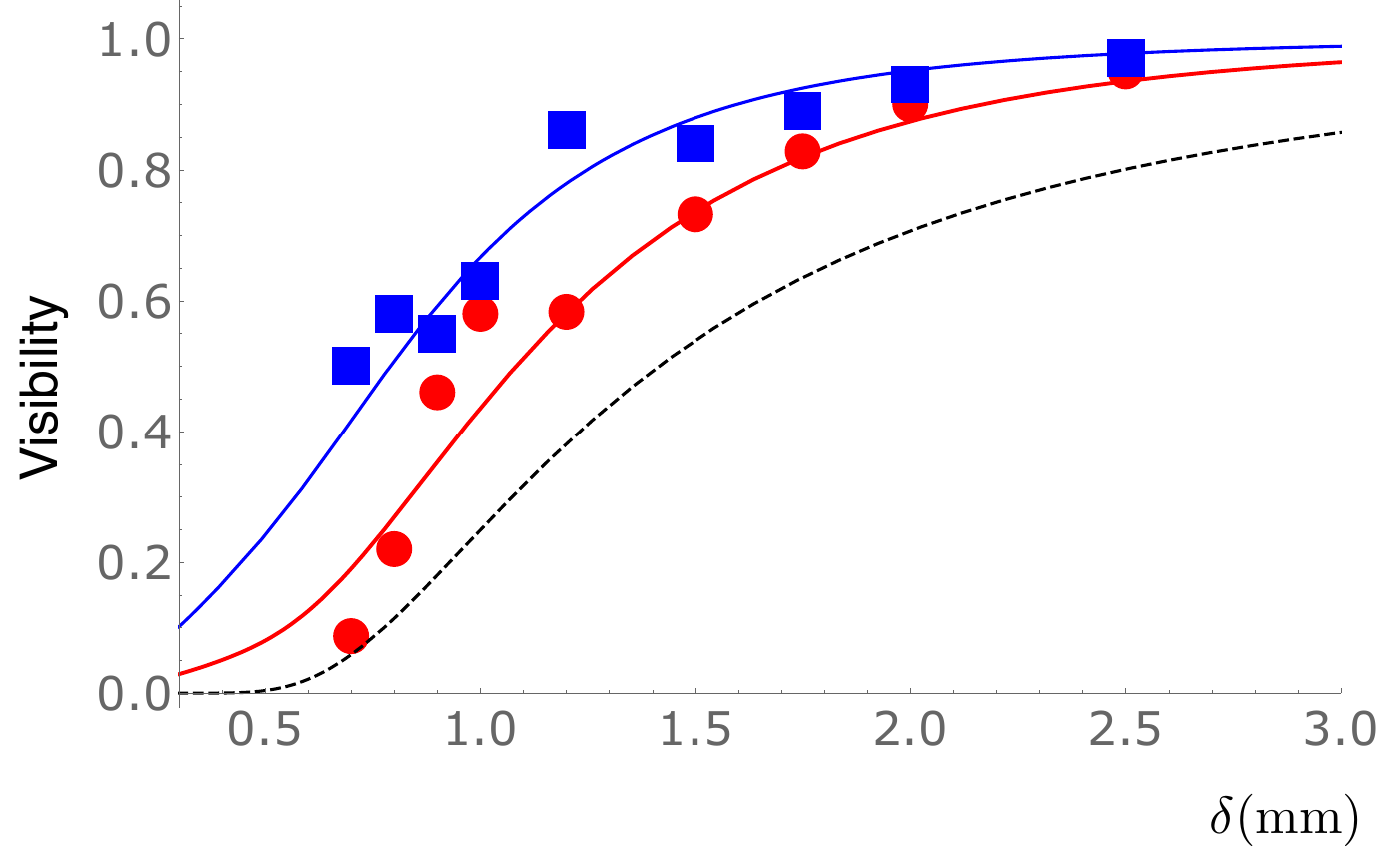}
 \caption{Visibility as a function of the transverse coherence length $\delta$ for normaized twist phase $\tau=0$. The black dashed curve is the TGSM visibility without coherent background given by \eqref{eq:visTGSM}. Red circles and blue squares are experimental data obtained with $L=300$ and $L=30$ video images, respectively. The associated red and blue curves are the visibility of the TGSM field with coherent background \eqref{eq:vistotal} with $\tau=0$. Error bars are smaller than the size of the symbols.}
 \label{fig:vist0}
 \end{figure}
 \begin{figure}
 \includegraphics[width=8.5cm]{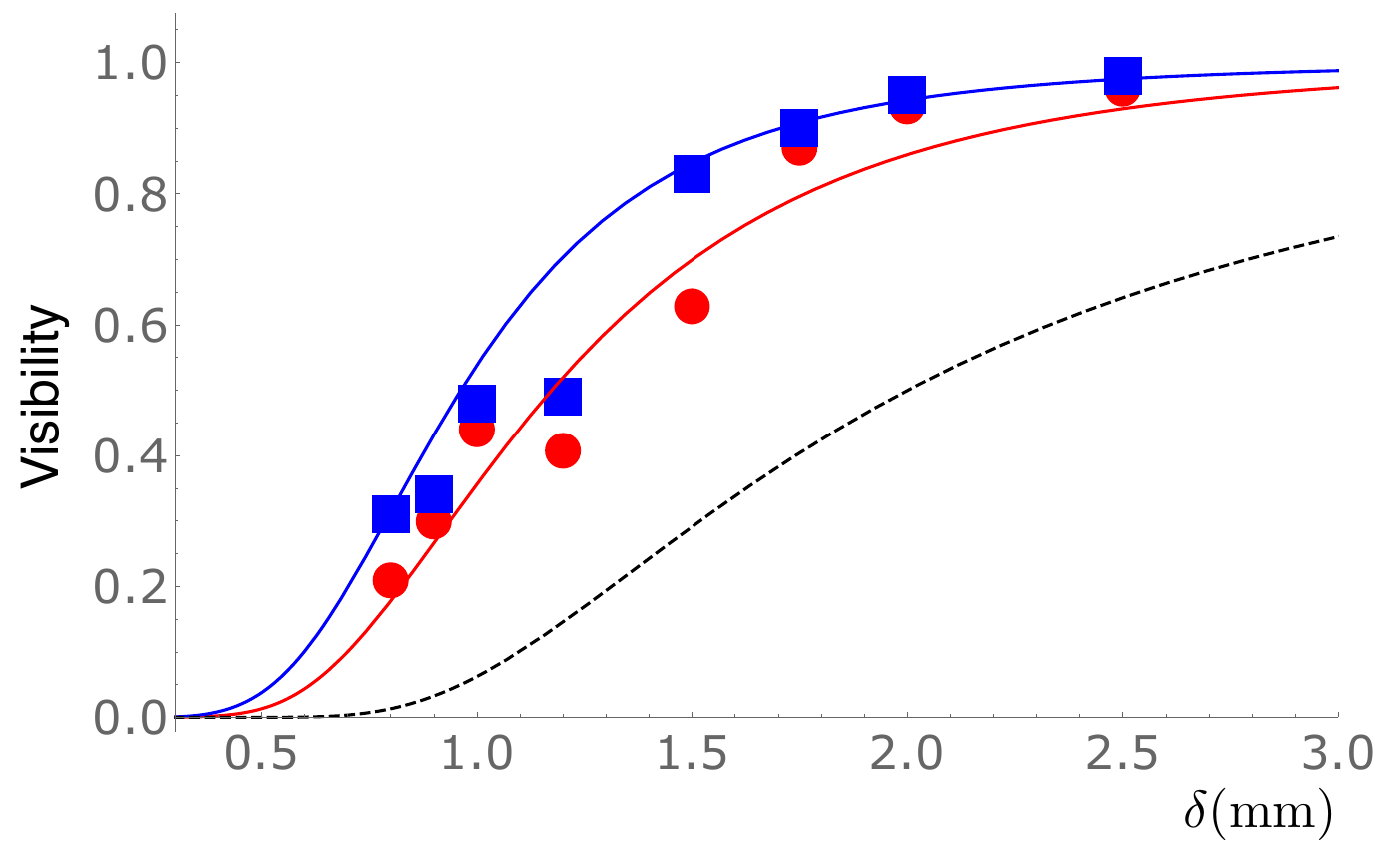}
 \caption{Visibility as a function of the transverse coherence length $\delta$ for normaized twist phase $\tau=1$. The black dashed curve is the TGSM visibility without coherent background given by \eqref{eq:visTGSM}. Red circles and blue squares are experimental data obtained with $L=300$ and $L=30$ video images, respectively. The associated red and blue curves are the visibility of the TGSM field with coherent background \eqref{eq:vistotal} with $\tau=1$. Error bars are smaller than the size of the symbols.}
 \label{fig:vist1}
 \end{figure}
 We perform the double slit experiment using the setup shown in Fig \ref{fig:setup} b), where the double-slit aperture is placed in the image plane of the SLM, created using spherical lenses with focal lengths $f_3=500$ mm and $f_4=150$ mm, giving a magnification factor of $0.333$. We use slits  separated by $500$ $\mu$m, and slit widths of $80$ $\mu$m. The effective slit dimensions, relative to the beam parameters at the plane of the SLM, are larger by a factor of three due to the imaging system, giving $d=250/0.33\approx758$ $\mu$m. We take images of the interference pattern for the entire duration of the SLM films, as described in the previous section. The marginal intensity distributions in the $x$-direction are obtained by summing the 2D images over the $y$-direction. Curve fits using Eq. \eqref{eq:int1} are used to estimate the visibility. A plot of the experimental data is shown in Figs. \ref{fig:vist0} and \ref{fig:vist1} for $\tau=0$ and $\tau=1$, respectively.  The red circles and blues squares correspond to TGSM beams obtained from $L=300$ and $L=30$ images. The solid curves will be described below. The black dashed curves give the theoretical prediction for the visibility using the TGSM expression \eqref{eq:visTGSM}.  Clearly, both figures show that the experimental data lies far from the theoretical curve, suggesting that the coherent background field must be taken into account to accurately describe the data.  
 \par
 Correcting for the coherent background, using Eq. \eqref{eq:wsum2}, the interference pattern should be a weighted sum of the interference patterns of the TGSM and the background field: $I_{tot}=(1-\Delta_L) I_{TG}+\Delta_L I_{coh}$. For the coherent component, we have
 \begin{equation}
 I_{coh}(x,\delta,\mu) = \gamma_{coh}(d,d) + \mathrm{Re}\left[\gamma_{coh}(d,-d) \right] \cos\left(\frac{2 d k x}{z} \right),
 \label{eq:int2}
 \end{equation}
 where we use the same shorthand notation for the coherent field as in Eq. \eqref{eq:int1}.  Using the CSD \eqref{eq:Wcoh}, one can check that $\mathrm{Re}\left[\gamma_{coh}(d,-d) \right] = \gamma_{coh}(d,d)$,  giving visibility $V_{coh}=1$.
 \par
 The overall visibility $\mathcal{V}_{total}$ of the entire field can then be calculated, giving
 \begin{equation}
 \mathcal{V}_{total} = \frac{(1-\Delta_L) \sqrt{ab} e^{-d^2 b}+4 \pi  \Delta_L}{(1-\Delta_L) e^{\frac{d^2 k^2 \mu^2}{a}} \sqrt{a b}+4 \pi   \Delta_L},
 \label{eq:vistotal}
 \end{equation}
 where $a$ is given in Eq. \eqref{eq:a} and here we define $b=a+2 \tau^2 \sigma^2$.
 Eq. \eqref{eq:vistotal} for $\tau=0$  is plotted as the solid curves in Fig. \ref{fig:vist0} using the estimated value of $\Delta_L=0.05$ ($\Delta_L=0.166$) for the $L=300$ ($L=30$) SLM images and the experimental parameters described above.  We can see that there is a much better correspondence with the experimental data. The  solid curves in Fig. \ref{fig:vist1} show plots of \eqref{eq:vistotal} for $\tau = 1$, and also shows much better agreement with the experimental data.  
 \par
\begin{figure}
 \includegraphics[width=8cm]{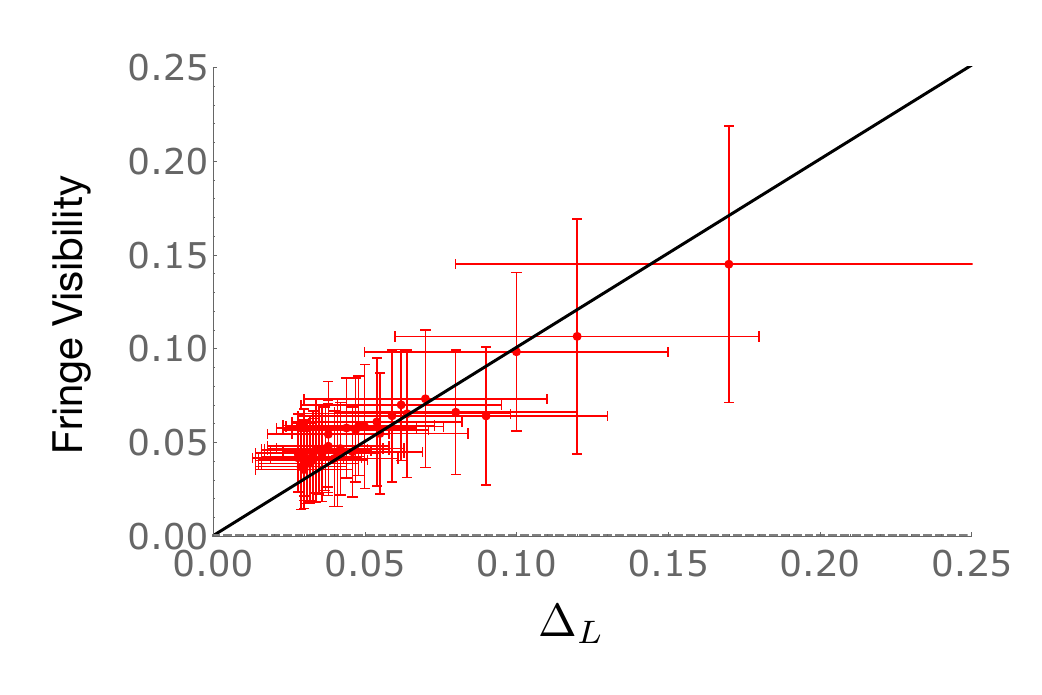}
 \caption{Visibility as a function of the coherence parameter $\Delta_L$. The red points are simulated experimental interference patterns with $\delta=0.4$mm, $\tau=0$ and other experimental parameters given in the text. The black curve corresponds to the theoretical model given by Eq. \eqref{eq:vistotal}.}
 \label{fig:VisVsDeltaL}
 \end{figure}

Our data shows that for a fixed coherence length $\delta$, a superior visibility is obtained with fewer video images (lower $L$), as a consequence of the larger coherent background. We further test this and our theoretical model by simulating experimental data. Figure \ref{fig:VisVsDeltaL} displays the visibility  as obtained from simulated interference patterns of the experiment for parameters $\delta=0.4$~mm and $\tau=0$. The mean and standard deviation of the visibility are calculated from a set of 30 random sequences of $L$ frames, where $L$ varied from 30 to 980. For each value of $L$, we obtain the mean and standard deviation of $\Delta_L$ by sampling $10^3$ values of the random phases and using Eq. \eqref{eq:DeltaL}.   The black solid curve is our theoretical prediction given by Eq. \eqref{eq:vistotal}. As can be observed, the values for the visibility are strongly correlated with those for $\Delta_L$ and are in agreement with our theoretical model, demonstrating the validity of the model and the importance of the residual coherence between the pseudo-modes.
 
 \section{Measuring Twist Phase}
\begin{figure}
 \includegraphics[width=8.5cm]{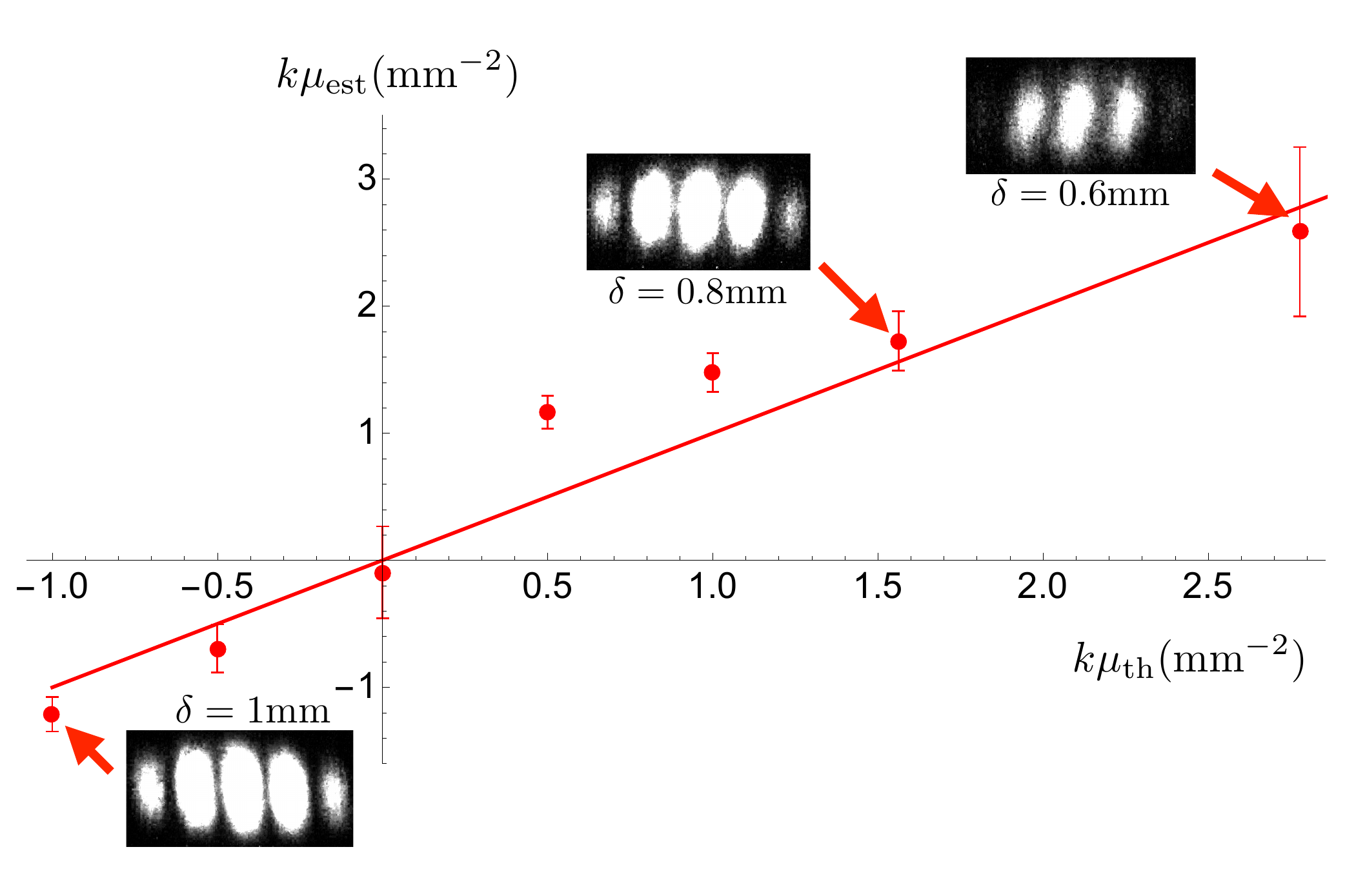}
 \caption{Comparison between theoretical and measured values of twist phase. Here $\mu_{th}$ is a theoretical value used to produce the TGSM beams, and $\mu_{est}$ is the  value estimated from the offset of the interference pattern. The red line is $\mu_{th}=\mu_{est}$. Experimental results are in agreement with the theory The insets show CCD camera images of the interference pattern obtained with synthesized TGSM beams. }
 \label{fig:twist}
 \end{figure}

 The square of the twist phase has observable effect on the far-field beam width as well as the double slit interference visibility discussed in the last subsections.  However, neither of these give an indication of the sign of the twist phase.  
 One method with which the twist phase has been measured experimentally is by observing the rotation of the beam as a function of the propagation distance \cite{tian20}. Here we provide a convenient alternative method, taking advantage of the fact that the 2D Young double slit interference pattern is twist-phase dependent. Let us consider a double slit aperture, with infinitesimal slits located at $x=\pm d$. After free-space propagation of length $z$ to the far-field, the intensity pattern is 
 \begin{equation}
 I(\boldsymbol{r}) \approx e^{-\frac{x^2}{s^2}}e^{-\frac{y^2}{t^2}}  \left[1 + e^{-\frac{2 a^2}{\delta^2}} \cos\left \{ dk \left ( \frac{x}{z} + 2 \mu y \right\} \right) \right]
 \label{eq:2dint}
 \end{equation}
 where $s$ and $t$ are two width parameters related to the optical system. One can see that the 2D interference pattern is shifted by a value that is proportional to $\mu$.  The insets in Fig. \ref{fig:twist} show examples of the shifted interferograms. From Eq. \eqref{eq:2dint}, we see that the central interference peak (cosine argument = 0) lies along a line in the transverse plane defined by $x=-2 \mu z y$. Thus, at positions $y=\pm h$, there is relative offset in the peak position given by $4 \mu h$, from which one can determine $\mu$. 
 \par
 We test our TGSM beams for different values of $k \mu$, shown in  Fig. \ref{fig:twist}. $\mu_{th}$ is the theoretical value used to produce the TGSM beam with the SLM, while $\mu_{est}$ is the value determined from the offset, calculated using five values of $h$ ranging between one and two standard deviations from $y=0$.  The solid line is $k \mu_{est}=k \mu_{th}$.  The experimental data agrees reasonably well with theory, validating the presence of twist phase in the TGSM beams as well this technique as a method for measuring twist phase (both magnitude and sign). Moreover, it should be possible to use this technique to measure twist phase in the correlations of photon pairs \cite{Hutter20,Hutter21}, either directly measuring a shift in the correlations as in \cite{gomes09a}, or using optical Fractional Fourier transforms \cite{tasca08}. 
 \section{Conclusion}
 In conclusion, we have analyzed a method to synthesize partially coherent twisted-Gaussian Schell Model beams from a coherent source, first demonstrated in Ref. \cite{tian20}. The technique uses a film sequence of images (300 in the present case) displayed on a spatial light modulator. When illuminated by a coherent Gaussian beam, a partially coherent beam is observed, provided the observation time is longer than the display time of the film. In the present work, the properties of the synthesized beams, such as near-field waist, far-field waist (divergence), transverse coherence length and twist phase were explored for a wide range of values of twist phase and coherence length. A theory was developed that includes the residual background that arises when a finite image sequence is used. For films composed of 300 images we observed that the residual coherence had null or negligible effect on the beam width in the near and far-field. However, it was observed that the residual coherence has observable effect on the effective transverse coherence length, when measured through Young double-slit interference.  This was well-described by the theoretical model.  The twist phase was measured using a novel technique based on 2D interferograms, where the twist phase produced a shift of the interference pattern. We found good agreement between theory and experiment.  Our results further validate the beam synthesis technique, and highlight the need to consider background coherence in certain instances.    
 \begin{acknowledgements}
This work was funded by the Chilean agencies Fondo Nacional de Desarrollo Cient\'{i}fico y Tecnol\'{o}gico (FONDECYT - DOI 501100002850) (1190901, 1190933, 1200266); National Agency of Research and Development (ANID) Millennium Science Initiative Program—ICN17-012; the Brazilian agencies
 Coordena\c c\~{a}o de Aperfei\c coamento de Pessoal de N\'\i vel Superior (CAPES DOI 501100002322), 
Funda\c c\~{a}o de Amparo \`{a} Pesquisa do Estado de Santa Catarina (FAPESC - DOI 501100005667),
Conselho Nacional de Desenvolvimento Cient\'{\i}fico e Tecnol\'ogico (CNPq - DOI 501100003593), 
Instituto Nacional de Ci\^encia e Tecnologia de Informa\c c\~ao Qu\^antica (INCT-IQ 465469/2014-0); the National Research Foundations of South Africa.  
 \end{acknowledgements}

 \bibliographystyle{apsrev}
 \bibliography{TGSM}


\end{document}